\documentstyle[12pt]{article}
\newcommand{\lesssim}
{\mathrel{\raisebox{-2.8pt}{\mbox{$\stackrel{\textstyle <}{\sim}$}}}}

\newcommand{\address}[1]{\centerline{\small\it #1}}
\newcommand{\draft}{}
\renewcommand{\title}[1]{\begin{center}%
{\large\bf #1}\end{center}\par\bigskip}
\renewcommand{\author}[1]{\centerline{#1}}
\renewcommand{\maketitle}{}
\newcommand{\pacs}[1]{}
\newcommand{\narrowtext}{}

\begin{document}
\draft
\title{From Quantum Dynamics to the Second Law
of Thermodynamics\footnote{
The first version  May 8, 2000.
}}
\author{Hal Tasaki$^*$}
\address{Department of Physics, Gakushuin University,
Mejiro, Toshima-ku, Tokyo 171, JAPAN}
\maketitle
\begin{abstract}
In quantum systems which satisfy the {\em hypothesis of 
equal weights for eigenstates}\/ \cite{Hal1}, 	
the maximum work principle
(for extremely slow and relatively fast operation)
is derived by using quantum dynamics alone.
This may be a crucial step in establishing a firm connection between 
macroscopic thermodynamics and microscopic quantum dynamics.
For special
models introduced in \cite{Hal1,Hal2}, the derivation of the maximum 
work principle can be executed without introducing any unproved 
assumptions.
\end{abstract}
\narrowtext

Although there is no doubt that the second law of thermodynamics 
is one of the most perfect and beautiful laws in physics, 
its connection 
to the rest of physics is still poorly understood.
It should be stressed that {\em equilibrium} statistical mechanics  
does {\em not}\/ lead to the second law.
The second law deals with 
transformations between two equilibrium states
caused by {\em any} macroscopically realizable processes 
which can be
far from equilibrium.
The second law sets sharp and highly nontrivial restrictions on 
the possibility of such transformations and on the energy exchange 
during the processes \cite{LY}.

A traditional approach toward
derivation of the second law, which goes back to
Boltzmann \cite{Boltzmann66}, 
has been to start from certain stochastic 
description of microscopic dynamics.
In the present note, we wish to concentrate on 
the possibility of deriving the 
second law from fully deterministic microscopic quantum
dynamics.
Such a link between quantum
mechanics and thermodynamics (if established) should not 
only provide a further basis for thermodynamics but also give an
indirect support to our belief (which can not been confirmed 
directly) that even macroscopic systems are governed by quantum
mechanics.
We shall here concentrate on the second law 
formulated as the {\em maximum work principle} 
(MWP) \cite{MWP}, 
and describe its derivation in quantum systems
which satisfy the conditions stated in \cite{Hal1} 
for two limiting situations of 
infinitely slow and relatively (but not infinitely)
fast operations.
Here we describe only basic ideas of the derivation, and leave 
details (which are simply technical and not difficult) 
to \cite{Hal2}.

\paragraph*{Basic setup and previous results:}
Let us start by recalling the general ideas and results in 
\cite{Hal1}, where we presented 
a scenario for deriving the canonical distribution from quantum 
dynamics, and an example in which such a derivation can be done 
without making any assumptions.
(See \cite{ref} for related attempts of deriving statistical physics 
using quantum dynamics.)
We consider 
an isolated quantum system 
which consists of a 
subsystem and a heat bath \cite{noteMC}.  
The subsystem alone 
is described by a Hamiltonian $H_{\rm S}$ which is
diagonalized as \( H_{\rm S}\Psi_{j}=\varepsilon_{j}\Psi_{j} \)
for \( j=1,2,\ldots,n \),
with \( \Vert \Psi_{j}\Vert=1 \) and 
\( \varepsilon_{j}<\varepsilon_{j+1} \).
Similarly the bath has a Hamiltonian \( H_{\rm B} \) which is 
diagonalized as
 \( H_{\rm B} \Gamma_{k}=B_{k}\Gamma_{k} \) 
 for \( k=1,2,\ldots,N \),
with \( \Vert \Gamma_{k}\Vert=1 \) and 
\( B_{k}\le B_{k+1} \).
We let \( \Omega_{\rm B}(B) \) a smooth function of \( B \)
such that \( \Omega_{\rm B}(B_{k})=k \)
({\em i.e.}\/,  \( \Omega_{\rm B}(B) \) is roughly 
the number of energy levels 
\( B_{k} \) with \( B_{k}\le B \)) and denote by 
\( \rho_{\rm B}(B)=d\Omega_{\rm B}(B)/dB \) the density of states of 
the bath.
Throughout the present note, we only consider a limited range of 
energy \( B \) in which these functions can be approximated as
\begin{equation}
	\Omega_{\rm B}(B)\simeq C_{1}+C_{2}e^{\beta B},
	\label{eq:OmegaB}
\end{equation}
and
\begin{equation}
	\rho_{\rm B}(B)\simeq C_{3}e^{\beta B},
	\label{eq:rhoB}
\end{equation}
with constants \( \beta \), 
\( C_{1} \), \( C_{2} \), and \( C_{3}=\beta C_{2} \).
Physically speaking, we are assuming that the bath is so large that 
its inverse temperature \( \beta \) does not vary when it exchanges 
energy (heat) with the subsystem during equilibration  and during
 operations.

The Hamiltonian for the whole system is
\begin{equation}
	H=(H_{\rm S}\otimes{\bf 1}_{\rm B})
	+({\bf 1}_{\rm S}\otimes H_{\rm B})
	+H_{\rm int},
	\label{eq:H}
\end{equation}
where \( {\bf 1}_{\rm S} \) and \( {\bf 1}_{\rm B} \) are
the identity operators,
and \( H_{\rm int} \) with \( \Vert H_{\rm int}\Vert=\lambda \) 
describes the interaction between the subsystem and the 
bath \cite{noteNorm}.
We assume that the bath is macroscopic and the interaction is weak in 
the sense that
\begin{equation}
	\Delta\varepsilon\gg\lambda\gg\Delta B,
	\label{eq:macro}
\end{equation}
where 
\( \Delta\varepsilon=\min_{j}\varepsilon_{j+1}-\varepsilon_{j} \)
and
\( \Delta B=\max_{k}B_{k+1}-B_{k} \) characterize the 
energy level spacings of the subsystem and the bath, respectively.

For \( \ell=1,\ldots,nN \), 
let us denote by \( \Phi_{\ell} \) the normalized eigenstate of the 
total Hamiltonian \( H \) with the eigenvalue \( E_{\ell} \).
We assume that the energy levels are nondegenerate and order them as
\( E_{\ell}<E_{\ell+1} \).
Let us expand the eigenstate as
\begin{equation}
	\Phi_{\ell}=\sum_{j,k}\varphi^{(\ell)}_{j,k}\,
	\Psi_{j}\otimes\Gamma_{k}.
	\label{eq:Phiexp}
\end{equation}
The {\em hypothesis of equal weights for eigenstates} proposed in 
\cite{Hal1} is that, for a {\em general}\/ interaction, 
the above coefficients \( \varphi^{(\ell)}_{j,k} \) 
satisfy
\begin{equation}
	|\varphi^{(\ell)}_{j,k}|^2
	\sim
	f(E-(\varepsilon_{j}+B_{k})),
	\label{eq:HEW}
\end{equation}
for {\em general}\/ \( \ell \) with 
\( E_{\ell} \) in a certain range \cite{note:general},
where the function \( f(x) \) has a single peak at \( x=0 \) and is 
negligible for \( |x|\ge C_{4}\lambda \),
where \( C_{4} \) is a constant.
The hypothesis looks  natural 
since, when \( \lambda=0 \), only \( (j,k) \) such that 
\( E-(\varepsilon_{j}+B_{k})=0 \) contribute to the 
expansion (\ref{eq:Phiexp}).
In \cite{Hal1}, we presented an artificial example in which this 
hypothesis can be established rigorously without any assumptions.
See \cite{Hal2} for a further (simpler) example.

Once accepting (\ref{eq:HEW}), 
it is easily observed that \cite{Hal1,Hal2}, 
for any operator \( A \) of the subsystem,
\begin{eqnarray}
	&&
	\langle\Phi_{\ell},(A\otimes{\bf 1}_{\rm B})
	\Phi_{\ell}\rangle
	\simeq
	\sum_{j,k}(A)_{j,j}|\varphi^{(\ell)}_{j,k}|^2
	\nonumber  \\
	&&
	\simeq
	\frac{\sum_{j}(A)_{j,j}\,
	\rho_{\rm B}(E_{\ell}-\varepsilon_{j})}
	{\sum_{j}\rho_{\rm B}(E_{\ell}-\varepsilon_{j})}
	\simeq
	\langle A\rangle^{\rm canonical}_{\beta},
	\label{eq:Acan1}
\end{eqnarray}
where the final estimate follows from (\ref{eq:rhoB}).
Here 
\( \langle\cdots\rangle^{\rm canonical}_{\beta} \) denotes
the canonical expectation at inverse temperature \( \beta \).
Furthermore it can be shown that for any initial state 
\begin{equation}
	\Phi(0)=\sum_{\ell}\gamma_{\ell}\,\Phi_{\ell},
	\label{eq:Phi0}
\end{equation}
with coefficients \( \gamma_{\ell} \) almost identically 
distributed for \( \ell \) such that 
\( |E_{\ell}-\bar{E}|\le\delta \),
for some \( \bar{E} \) and a  constant \( \delta \) satisfying
\( \Delta B\ll\delta \ll \Delta\varepsilon\),
one has 
\begin{equation}
	\langle\Phi(t),(A\otimes{\bf 1}_{\rm B})
	\Phi(t)\rangle
	\simeq
	\langle A\rangle^{\rm canonical}_{\beta},
	\label{eq:canA2}
\end{equation}
for sufficiently large and typical \( t \),
where \( \Phi(t)=e^{-iHt}\Phi(0) \) is the state at time \( t \).
We have therefore shown (under the hypothesis about the 
structure of eigenstates) that quantum dynamics alone
brings the system into the canonical distribution.

\paragraph*{External operation and work:}
We wish to treat a situation typical in thermodynamics, 
where an external agent performs an operation to the subsystem 
(e,g, moving a piston attached to a cylinder)
leaving the bath untouched .
We model the operation as a change of the Hamiltonian of the subsystem.
More precisely,  the Hamiltonian for the subsystem is \( H_{\rm S}(t) \)
with  \( H_{\rm S}(t)=H_{\rm S} \) for \( t\le t_{0} \) and
  \( H_{\rm S}(t)=H'_{\rm S} \) for \( t\ge t_{0}+\tau \).
The operation takes place between \( t_{0} \) and \( t_{0}+\tau \),
and the Hamiltonian is constant otherwise.
We denote by \( \varepsilon_{j'}' \) the eigenvalues
of \( H'_{\rm S} \).

Let \( \Phi(0) \) be the initial state as in (\ref{eq:Phi0}), and
let \( \Phi(t) \) be its time evolution determined by the 
time-dependent Hamiltonian
\( H(t)=(H_{\rm S}(t)\otimes{\bf 1}_{\rm B})
+({\bf 1}_{\rm S}\otimes H_{\rm B})+H_{\rm int} \).
We assume that \( t_{0} \) is chosen sufficiently large 
so that  \( \Phi(t_{0}) \), which is the state right before the 
operation, describes 
the thermal equilibrium in the sense of (\ref{eq:canA2}).
When  \( t \) becomes sufficiently large,
the state \( \Phi(t) \) is expected to
reach the new equilibrium after the operation
 \cite{note:pure}.

From the energy conservation law, one finds that the work done by the 
subsystem to the external agent \cite{note:classical} is
\begin{equation}
	W=\langle\Phi(t_{0}),H\,\Phi(t_{0})\rangle
	-\langle\Phi(t_{0}+\tau),H'\,\Phi(t_{0}+\tau)\rangle,
	\label{eq:W}
\end{equation}
where \( H' \) denotes the total Hamiltonian for \( t\ge t_{0}+\tau \).
The {\em maximum work principle}\/ 
(MWP)
states that the above work satisfies the inequality
\begin{equation}
	W\le F(\beta)-F'(\beta),
	\label{eq:MWP}
\end{equation}
for any operations, and the equality holds if the operation is done 
infinitely slowly.
Here \( F(\beta)=-\beta^{-1}\log\sum_{j}e^{-\beta\varepsilon_{j}} \)
and  \( F'(\beta)=-\beta^{-1}\log\sum_{j}e^{-\beta\varepsilon'_{j}} \)
are the free energies of the subsystem
before and after the operation, respectively.

\paragraph*{Slow operation:}
We first consider infinitely slow operation realized in the 
\( \tau\to\infty \) limit, which corresponds to quasi-static 
operations in thermodynamics.
In this limit, time evolution of the state \( \Phi(t) \) is completely 
determined by the adiabatic theorem in quantum mechanics 
\cite{note:adiabatic} if we 
assume that the Hamiltonian \( H(t) \) has no degenerate eigenstates 
for any \( t \).
If one starts from one of the eigenstates of \( H \), the 
time evolution of the state exactly traces  the 
corresponding eigenstate of \( H(t) \) during 
the operation.
Thus if we start from \( \Phi(0) \) of the form (\ref{eq:Phi0}),
the state right after the operation is written as
\( \Phi(t_{0}+\tau)=\sum_{\ell}
\gamma_{\ell}\,\theta_{\ell}\,\Phi'_{\ell} \),
with \( |\theta_{\ell}|=1 \).
Here \( \Phi'_{\ell} \) is the 
eigenstate of \( H' \) with the eigenvalue \( E'_{\ell} \),
where the energy levels are 
again ordered as \( E'_{\ell}<E'_{\ell+1} \).

In order to estimate the work done by the subsystem, 
we introduce the index \( \bar{\ell} \) such that
\( E_{\bar{\ell}}=\bar{E} \), where \( \bar{E} \) is (roughly)
 the mean energy of the state \( \Phi(t) \) before the operation.
The mean energy after the operations is simply given by 
\( \bar{E}'_{\rm slow}=E'_{\bar{\ell}} \).
Therefore the energies \( \bar{E} \) and  
\( \bar{E}'_{\rm slow} \) 
are related by 
\begin{equation}
	\Omega(H\le\bar{E})=\Omega(H'\le\bar{E}'_{\rm slow}),
	\label{eq:OO1}
\end{equation}
where \( \Omega(A\le a) \) denotes the number of eigenstates of a 
hermitian matrix \( A \) with eigenvalues less than or equal 
to \( a \) \cite{note:S1}.

Since \( \Vert H_{\rm int}\Vert=\lambda \), we can neglect 
\( H_{\rm int} \) in (\ref{eq:OO1}) to get 
\begin{equation}
	\Omega(H_{\rm S}+H_{\rm B}\le\bar{E})
	\simeq
	\Omega(H'_{\rm S}+H_{\rm B}\le\bar{E}'_{\rm slow}),
	\label{eq:OO2}
\end{equation}
with errors of \( O(\lambda) \) in the energies
\cite{note:minimax}.
By treating the energy levels of \( H_{\rm S} \) and  \( H'_{\rm S} \) 
explicitly, we can rewrite (\ref{eq:OO2}) as
\begin{equation}
	\sum_{j=1}^n\Omega_{\rm B}(\bar{E}-\varepsilon_{j})
	\simeq
	\sum_{j=1}^n\Omega_{\rm B}(\bar{E}'_{\rm slow}-\varepsilon'_{j}).
	\label{eq:OO3}
\end{equation}
By using (\ref{eq:OmegaB}),
the relation (\ref{eq:OO3}) immediately implies
\begin{equation}
	W_{\rm slow}\equiv\bar{E}-\bar{E}'_{\rm slow}
	\simeq F(\beta)-F'(\beta),
	\label{eq:Wslow}
\end{equation}
which is the equality corresponding to the desired MWP (\ref{eq:MWP}).

\paragraph*{Fast operation:}
Next we consider the opposite situation where the operation is 
executed quickly.
We assume that the duration of the operation 
satisfies \( \tau\ll\lambda^{-1} \).
In other words, the operation is done so quickly that 
the subsystem and the bath essentially do not exchange energy (heat)
during the operation.
The exchange of heat takes place in the equilibration process 
after the operation.

Since we have chosen \( t_{0} \) so that \( \Phi(t_{0}) \) describes 
the equilibrium, it can be expanded as
\begin{equation}
	\Phi(t_{0})=\sum_{j,k}\xi_{j,k}\Psi_{j}\otimes\Gamma_{k},
	\label{eq:Phitexp}
\end{equation}
where the coefficients \( \xi_{j,k} \) satisfy the 
hypothesis of equal weight (\ref{eq:HEW}) just as 
\( \varphi^{(\ell)}_{j,k} \).

Let us consider the time evolution during the operation.
From the assumption of quick operation, 
the state of the bath essentially remains unchanged 
while that of the subsystem changes according to a 
unitary transformation
\( \Psi_{j}\to\sum_{j'}U_{jj'}\Psi'_{j} \).
Here we diagonalized \( H'_{\rm S} \) as 
\( H'_{\rm S}\Psi'_{j}=\varepsilon'_{j}\Psi'_{j} \)
with \( \varepsilon'_{j}<\varepsilon'_{j+1} \).

Then the state immediately after the operation is
\begin{equation}
	\Phi(t_{0}+\tau)
	\simeq
	\sum_{j,j',k}\xi_{j,k}U_{jj'}\Psi_{j'}\otimes\Gamma_{k}.
	\label{eq:Phiafter}
\end{equation}
We can evaluate the energy expectation value of this state
as in (\ref{eq:Acan1}) to get
\begin{eqnarray}
	&&
	\bar{E}'_{\rm fast}\equiv
	\langle\Phi(t_{0}+\tau),H'\,\Phi(t_{0}+\tau)\rangle
	\nonumber\\
	&&
	=
	\sum_{j',k}|\sum_{j}\xi_{j,k}U_{jj'}|^2
	2(B_{k}+\varepsilon'_{j'}+O(\lambda))
	\nonumber\\
	&&
	\simeq
	\frac{\sum_{j,j'}\rho_{\rm B}(\bar{E}-\varepsilon_{j})
	|U_{jj'}|^2(\bar{E}-\varepsilon_{j}+\varepsilon'_{j'})}
	{\sum_{j}\rho_{\rm B}(\bar{E}-\varepsilon_{j})}
	\nonumber\\
	&&
	\ge
	\frac{\sum_{j}\rho_{\rm B}(\bar{E}-\varepsilon_{j})
	(\bar{E}-\varepsilon_{j}+\varepsilon'_{j})}
	{\sum_{j}\rho_{\rm B}(\bar{E}-\varepsilon_{j})},
	\label{eq:Efast}
\end{eqnarray}
where we used (\ref{eq:HEW}) to get the third line.
The final 
inequality follows \cite{note:ineq} 
by noting that \( \varepsilon'_{j'} \) 
is increasing in \( j' \) while 
\( \rho_{\rm B}(\bar{E}-\varepsilon_{j}) \) is decreasing in \( j \),
and \( \sum_{j}|U_{jj'}|^2=\sum_{j'}|U_{jj'}|^2=1 \).
On the other hand,
sine \( \Omega_{\rm B}(B) \) is convex in \( B \),
(\ref{eq:OO3}) implies that
\begin{eqnarray}
	&&
	\sum_{j=1}^n\Omega_{\rm B}(\bar{E}-\varepsilon_{j})
	\simeq
	\sum_{j=1}^n\Omega_{\rm B}(\bar{E}'_{\rm slow}-\varepsilon'_{j})
	\nonumber\\
	&&
	=
	\sum_{j=1}^n\Omega_{\rm B}(\bar{E}-\varepsilon_{j}
	+(\bar{E}'_{\rm slow}-\bar{E}-\varepsilon'_{j}+\varepsilon_{j}))
	\nonumber\\
	&&
	\ge
	\sum_{j=1}^n\Omega_{\rm B}(\bar{E}-\varepsilon_{j})
	+\sum_{j=1}^n
	(\bar{E}'_{\rm slow}-\bar{E}-\varepsilon'_{j}+\varepsilon_{j})
	\rho_{\rm B}(\bar{E}-\varepsilon_{j}),
	\label{eq:OO4}
\end{eqnarray}
and hence
\begin{equation}
	\bar{E}'_{\rm slow}\lesssim
	\frac{\sum_{j}\rho_{\rm B}(\bar{E}-\varepsilon_{j})
	(\bar{E}-\varepsilon_{j}+\varepsilon'_{j})}
	{\sum_{j}\rho_{\rm B}(\bar{E}-\varepsilon_{j})}.
	\label{eq:Eslow<}
\end{equation}
Then from (\ref{eq:Efast}), we find
\( \bar{E}'_{\rm slow}\lesssim\bar{E}'_{\rm fast} \) \cite{note:S2},
and by recalling (\ref{eq:Wslow}), we find
\begin{equation}
	W_{\rm fast}\equiv\bar{E}-\bar{E}'_{\rm fast}
	\lesssim
	W_{\rm slow}\simeq F(\beta)-F'(\beta),
	\label{eq:Wfast}
\end{equation}
which is the desired MWP.

\paragraph*{Discussions:}
We have derived the MWP for infinitely slow and 
relatively (but not infinitely)
fast operations by using  quantum dynamics and our {\em 
hypothesis of equal weights for eigenstates}\/.
Note that since
the hypothesis has been proved for some 
models \cite{Hal1,Hal2}, we now have derived 
rigorously the (parts of the) second law of thermodynamics in  
concrete quantum mechanical models \cite{note:rigorous}.
As we have discussed 
in \cite{Hal1,Hal2}, we believe that our hypothesis 
is valid in a rather general class of quantum systems.

Among many questions to be discussed, 
let us address two particularly important ones.
The first natural question is
whether the MWP can be 
derived for general operations which are neither extremely slow nor 
very quick.
A naive perturbative estimate 
around the \( \tau\to\infty \) limit suggests 
the validity of the MWP, 
but to construct rigorous estimate 
from such a heuristic calculation seems quite difficult.
A rigorous analysis of general operations seems formidably 
difficult since we 
have almost no ways of treating general time evolution in quantum 
systems with time dependent Hamiltonians.
Moreover although unquestionable success of thermodynamics may seem to 
suggest the universal validity of the MWP, one should note that 
in experiments one only encounters operations which are realized 
as motions of macroscopic objects.
There is a possibility that a very carefully designed time-dependent 
Hamiltonian \( H_{\rm S}(t) \) leads to a time evolution which 
violates the MWP.
If this is the case, 
all that we can hope to prove is the validity of the 
MWP for a limited class of operations which are 
``macroscopically realizable.''
For the moment, we have no idea about what criteria should we use to 
distinguish such operations.

The second question is whether our result applies to realistic 
situations where one applies many operations repeatedly 
to the subsystem.
To answer this, suppose that we start from an initial state 
(\ref{eq:Phi0}) where \( \gamma_{\ell} \) is nonvanishing only for 
\( \ell \) such that \( |\bar{E}-E_{\ell}|\le\delta \).
After a general operation, 
we end up with a similar state, but with a different mean energy
\( \bar{E}' \)
and the energy range \( \delta' \) which is in general strictly 
larger than than the initial \( \delta \).
Therefore if we repeat general operations sufficiently 
many times, 
the range \( \delta \) becomes large and may violate the required 
condition \( \delta\ll\Delta\varepsilon \).
Therefore, technically speaking, although 
we {\em can} use the present result 
as long as the number of operations does not exceed a certain limit 
(which limit depends on the initial state and the nature of the 
operations),
there is no hope of dealing with indefinitely many operations.
We still do not know if this limitation contradicts with our 
experiences that the second law of thermodynamics has been confirmed 
in repeated experiments \cite{note:repeat}.

\bigskip
It is a pleasure to thank
Tohru Koma,
Yoshi Oono,
and
Shin-ichi Sasa
for stimulating discussions on various related topics.


\begin{thebibliography}{10}
\bibitem{noteMC}
One can also try recovering thermodynamics (in the 
\( (U,X) \)-representation as in \cite{LY}) from a single isolated 
quantum system.
We feel, however, that an approach using the bath and the subsystem
is less involved both conceptually and technically.
\bibitem{note:general}
Of course it is a highly debatable question to decide which quantum 
systems and which quantum states are ``general.''
\bibitem{note:pure}
As in \cite{Hal1}, we here concentrate on the situation where 
\( \Phi(t) \) is a pure state,
but to treat mixed states is a trivial extension.
\bibitem{note:classical}
We are implicitly assuming that the external agent is ``classical'', 
and the difference \( W \) between the
two quantum mechanical expectation
values can be regarded as the classical work.
\bibitem{note:S1}
For a (pure) state of the whole system whose energy is concentrated 
at \( \bar{E} \), we can define its Boltzmann entropy by
\( S=k\log\Omega(H\le\bar{E}) \).
Then (\ref{eq:OO1}) states that the entropy of a closed system 
remains unchanged during a slow operation, and this is true for an 
arbitrary slow operation on the whole system.
Such an implication of the adiabatic theorem has been of course 
wellknown.
\bibitem{note:minimax}
More precisely it is well known  \cite{Hal2} that Weyl's minimax 
principle implies the bounds
\( \Omega(H_0\le\bar{E}-\lambda)\le 
\Omega(H\le\bar{E})\le
\Omega(H_0\le\bar{E}+\lambda) \)
(and the corresponding bounds for \( H' \)), where
\( H_{0}=H-H_{\rm int} \).
Throughout the present note,
approximate equalities (\( \simeq \)) and inequalities (\( \lesssim \))
should be regarded as rigorous equalities and inequalities, respectively,
but with small errors of \( O(\lambda) \) or \( O(\delta) \).
\bibitem{note:ineq} 
Let \( f_{i} \), \( g_{i} \) be nondecreasing in \( i \),
and \( \alpha_{ij}\ge0 \) satisfy 
\( \sum_{i}\alpha_{ij} = \sum_{j}\alpha_{ij} = 1 \).
Then by expanding the left-hand side of the trivial inequality
\( \sum_{i,j}(f_{i}-f_{j})\alpha_{ij}(g_{i}-g_{j})\ge0 \)
one gets 
\( \sum_{i,j}f_{i}\alpha_{ij}g_{j}\le\sum_{i}f_{i}g_{i} \).
\bibitem{note:S2}
As in \cite{note:S1}, this implies the corresponding inequality 
\( S\le S' \) for the Boltzmann entropy for the whole closed system.
Unlike the corresponding equality \cite{note:S1}, however, 
the same inequality for {\em arbitrary} fast (or instantaneous)
operations on the whole system is not valid.
(One can easily write down counterexamples.)
We expect the inequality to
be valid for generic macroscopic operations.
\bibitem{note:rigorous}
To prove  (\ref{eq:Wslow}) rigorously in the model of \cite{Hal1}, 
we have to 
let \( H_{\rm int} \) depend on \( t \).
\bibitem{note:repeat}
One may argue that, in reality, the initial 
energy range \( \delta \) happened to be extraordinarily 
small, and 
realistic operations (which are not too fast) do not increase the 
range \( \delta \) too much.
One may also argue, however, that we are not allowed to rely 
on initial condition too much, and must consider some mechanisms (like 
``decoherence''?) which ``refreshes'' quantum states by narrowing the 
range \( \delta \).
\end{thebibliography}
\end{document}